\documentclass[letterpaper,twocolumn,10pt]{article}
\usepackage[letterpaper,left=0.75in,right=0.75in,top=1in,bottom=1in,columnsep=0.33in]{geometry}
\usepackage{mathptmx}
\usepackage[T1]{fontenc}
\usepackage[utf8]{inputenc}
\usepackage{pslatex}
\usepackage[kerning,spacing]{microtype}
\usepackage{flushend}
\usepackage{cite,url,xcolor}
\usepackage{hyperref}
\usepackage{amsmath,amssymb,graphicx,booktabs,tabularx,array}
\usepackage{enumitem,caption,placeins,multirow}
\hypersetup{hidelinks,pdfauthor={Fei Li, Song Liu, Shiqiang Nie, Jinyu Wang, Weiguo Wu},pdfsubject={arXiv preprint},
  pdftitle={SpecStream: Co-Optimizing KV Cache Management and Model Scheduling for Memory-Constrained Speculative Decoding}}
\setlist[itemize]{leftmargin=*,topsep=3pt,itemsep=3pt,parsep=0pt}
\makeatletter
\g@addto@macro\normalsize{%
  \abovedisplayskip=6pt plus 1pt minus 1pt
  \belowdisplayskip=5pt plus 1pt minus 1pt
  \abovedisplayshortskip=3pt plus 1pt
  \belowdisplayshortskip=4pt plus 1pt minus 1pt
}
\newcommand{\nativepdf}[2][]{\begingroup\@ifundefined{Gread@transgrouptrue}{}{\let\Gread@transgrouptrue\Gread@transgroupfalse}\includegraphics[#1]{#2}\endgroup}
\renewcommand\section{\@startsection{section}{1}{\z@}{-2.4ex plus-.6ex minus-.2ex}{1.2ex plus.2ex}{\reset@font\large\bfseries\raggedright\hyphenpenalty=10000}}
\renewcommand\subsection{\@startsection{subsection}{2}{\z@}{-2.0ex plus-.5ex minus-.2ex}{1.0ex plus.2ex}{\reset@font\large\bfseries\raggedright\hyphenpenalty=10000}}
\renewcommand\subsubsection{\@startsection{subsubsection}{3}{\z@}{-1.8ex plus-.4ex minus-.2ex}{.8ex plus.2ex}{\reset@font\normalsize\bfseries\raggedright\hyphenpenalty=10000}}
\def\@maketitle{%
  \newpage
  \begin{center}
    {\Large\bfseries\@title\par}
    \vskip 0.8em
    {\normalsize\@author\par}
  \end{center}
  \vskip 0.5em
}
\renewenvironment{abstract}{%
  \begin{center}{\large\bfseries\abstractname\vspace{-.5em}\vspace{\z@}}\end{center}%
}{}
\makeatother
\date{}
\author{%
  {\large Fei Li\quad Song Liu\quad Shiqiang Nie\quad Jinyu Wang\quad Weiguo Wu}\\[0.45em]
  School of Computer Science and Technology, Xi'an Jiaotong University\\
  Xi'an, China\\[0.35em]
  {\small
    \href{mailto:lifei@stu.xjtu.edu.cn}{\nolinkurl{lifei@stu.xjtu.edu.cn}}\quad
    \href{mailto:liusong@mail.xjtu.edu.cn}{\nolinkurl{liusong@mail.xjtu.edu.cn}}\quad
    \href{mailto:shiqiang.nie@xjtu.edu.cn}{\nolinkurl{shiqiang.nie@xjtu.edu.cn}}}\\
  {\small
    \href{mailto:jinyu.wang@xjtu.edu.cn}{\nolinkurl{jinyu.wang@xjtu.edu.cn}}\quad
    \href{mailto:wgwu@xjtu.edu.cn}{\nolinkurl{wgwu@xjtu.edu.cn}}}%
}

\title{Resource-Efficient Speculative Decoding for Long-Context LLM Serving}

\begin{document}
\maketitle
\begin{abstract}

Speculative decoding reduces sequential Target model calls by verifying multiple tokens from the Draft model in parallel. Yet KV Cache growth limits long-context serving under constrained GPU memory. Offloading KV to CPU memory relieves this pressure. However, existing offloading schemes restore the full KV history before attention and fail to fully exploit the benefits of KV sharing across queries within a verification round. Existing parallel speculative decoding methods also overlook idle GPU compute capacity while memory-bandwidth-bound Target verification waits for historical KV. We present SpecStream, a speculative decoding system that begins verification without waiting for the full KV history to be restored and exploits compute bubbles during KV transfers for concurrent drafting on the same GPUs. It offloads only Target-committed history, keeping candidate rollback local to the GPUs. Each streamed KV chunk serves all queries in the round, with online softmax preserving full attention. Target-priority scheduling controls Draft concurrent execution under resource limits to constrain interference with Target verification. Experiments show that SpecStream maintains task quality close to SGLang speculative decoding while supporting more concurrent requests under limited GPU memory. Across different datasets, it achieves average throughput speedups of 1.41\ensuremath{\times} and 1.32\ensuremath{\times} over the offloading baseline for Qwen3 and InternLM2.5, respectively. Compared with parallel speculative decoding on separate Target and Draft GPUs, SpecStream improves output throughput per GPU by an average of 55.4\%.

\end{abstract}

\section{Introduction}\label{sec:1}

Long-context generation imposes a growing memory burden on large language model (LLM) serving. Models typically generate output autoregressively, one token at a time, and each new position attends to the existing context \cite{ref40}. To avoid recomputing attention Key-Value (KV) for earlier positions, the system retains them in a KV Cache \cite{ref01}. Model weights remain largely fixed during inference, whereas the KV Cache grows as sequences advance and consumes GPU memory across active requests. Prefill processes input positions in parallel. Decoding, by contrast, advances only a few tokens per forward pass while repeatedly accessing historical KV states. Efficient long-context serving therefore depends not only on compute speed but also on the ability to retain and access this history efficiently. GPU memory capacity and data movement are central constraints \cite{ref02}.

Speculative decoding \cite{ref03,ref04} accelerates token generation by reducing sequential calls to the Target model, which determines the final output. A lightweight Draft model first generates several candidates. The Target then verifies them in a single forward pass, potentially advancing multiple output positions at once. Candidates become committed output only after acceptance and correction. Recent work improves candidate quality and verification efficiency through trained drafters \cite{ref06}, multi-head proposals \cite{ref05}, and tree-structured verification \cite{ref07}. Fewer Target calls, however, do not remove the dependence of each verification round on historical KV states. The weights and KV Caches of both models, candidate states, and execution buffers all compete for GPU memory. Speculative decoding thus shortens the sequential execution path but may also reach the GPU memory limit sooner.

Offloading the KV history to larger CPU memory provides a direct way to relieve GPU memory pressure. Existing systems improve KV management through tiered storage \cite{ref02}, cache reuse \cite{ref08}, and selective retrieval \cite{ref09}. In speculative decoding, migration must also account for the Target's commit state. Otherwise, the system may transfer and maintain KV states for candidates that are later rejected. Restricting migration to committed history avoids this wasted work, but verification still has to wait for data. If every round restores the full history before computation begins, transfer time grows with context length and extends the critical path of Target verification. Offloading therefore requires more than a change in storage location. The Target must be able to compute with incoming KV states promptly and reuse them effectively.

KV access in speculative decoding differs from that in single-token autoregressive decoding. The queries in a Target verification round have different causal visibility within the candidate region but share the same earlier committed KV history. A single history fetch can therefore serve the entire verification query group, amortizing its transfer cost over the tokens actually committed in that round. Optimizing historical KV access therefore requires more than faster transfers; it must also account for the useful output each transfer supports. In addition, KV offloading paths in existing speculative decoding systems \cite{ref32,ref33} still defer attention computation until the current layer's full KV history is available on the GPU, preventing already transferred chunks from being processed as they arrive.

Waiting for historical KV leaves GPU compute capacity idle during memory-bandwidth-bound Target verification. With historical KV offloaded to CPU memory, verification requires host-to-device (H2D) transfers, which can further extend these waits. Even with partial overlap between transfers and computation, the Target can still stall if the current computation finishes before the next KV chunk arrives, creating compute bubbles. MineDraft \cite{ref10}, SPECTRE \cite{ref11}, and SwiftSpec \cite{ref12} hide part of the drafting latency by overlapping candidate generation with Target verification through cross-GPU parallelism or pipelined scheduling. These designs, however, do not directly exploit these bubbles to advance the Draft on the Target GPUs.


These observations motivate SpecStream, which jointly optimizes historical KV access and speculative execution. SpecStream uses the Target commit boundary to constrain history migration, keeping candidate rollback local to the GPUs. During Target verification, SpecStream combines chunked transfers, KV sharing across queries, and online softmax to let computation advance as history chunks arrive, without waiting for the current layer’s full history to be restored, while preserving full causal attention. Dynamic verification width further balances the benefits of history reuse against additional computation. Building on this design, SpecStream schedules Draft co-execution on the same GPUs under a Target latency budget, using spare compute capacity during transfers to advance candidate generation. This reduces the drafting latency exposed on the critical path and the reliance on additional GPUs. The contributions are summarized as follows:

\begin{itemize}
\item We propose commit-aware rollback-free KV tiering. SpecStream uses the Target commit boundary to constrain migration, offloading irrevocably committed older history to CPU memory while retaining recent KV states and unverified candidates on the GPU. History migration and speculative rollback operate on disjoint state regions. Batched migration also reduces fragmented transfers.

\item We propose KV streaming for multi-query verification. SpecStream divides the offloaded KV into contiguous chunks. Each chunk serves every query in the current verification round after it reaches the GPU, and online softmax combines the history chunks with recent GPU-resident KV states to compute full causal attention. Grouped transfers, double buffering, and cross-layer prefetching overlap data movement with computation. Cohort batching aggregates compatible tasks within a batch, while the verification width adapts to transfer cost, candidate acceptance, and compute overhead to improve the efficiency of KV history transfers.

\item We propose Target-priority intra-GPU co-execution. SpecStream combines offline measurements with runtime observations to estimate available execution windows under the Texture Processing Cluster (TPC) allocation selected for the Draft. It schedules subsequent candidate generation one token at a time. Completion feedback updates execution estimates and admission decisions for the next step, allowing the Draft to use spare compute capacity during KV history transfers while controlling interference with Target computation.

\item We implement SpecStream in SGLang and evaluate it with Qwen3 and InternLM2.5. Across three different datasets, SpecStream supports more concurrent requests under limited GPU memory than methods without offloading and achieves average throughput speedups of 1.41\ensuremath{\times} and 1.32\ensuremath{\times}, respectively, over the offloading baseline. Its shared-GPU execution improves output throughput per GPU by an average of 55.4\% compared with speculative decoding that places the two models on separate GPUs.

\end{itemize}

\section{Background}\label{sec:2}

\subsection{Speculative Decoding}\label{sec:2.1}

Standard autoregressive decoding generates one new token per round. Speculative decoding introduces a computationally cheaper Draft model so that a single Target verification pass can advance multiple output positions \cite{ref03,ref04}. At the start of round $r$, let the length of the confirmed output prefix be ${N}_{r}$, and denote it by ${x}_{1:{N}_{r}}$. Let $q$ denote the verification width. Conditioned on this prefix, the Draft generates $q$ consecutive candidate tokens as follows:
\begin{equation}
\widetilde{x}_{N_r+i}\sim p_{\mathrm D}(\cdot\mid x_{1:N_r},\widetilde{x}_{N_r+1:N_r+i-1}),\quad 1\le i\le q .
\label{eq:1}
\end{equation}
Here, $p_{\mathrm D}$ is the Draft’s conditional next-token distribution.

A single Target forward pass produces the Target distributions for the candidate positions. The acceptance and correction rules then determine the committed output. Let ${U}_{r}\left( q\right)$ denote the number of output tokens committed in this round. Ignoring changes in batching and assuming sequential execution of drafting and verification, the mean execution cost per committed token is approximated by
\begin{equation}
{\overline{T}}_{\mathrm{SD}}\left( q\right)=\frac{{T}_{\mathrm{D}}\left( q\right)+{T}_{\mathrm{V}}\left( q\right)+{T}_{\mathrm{R}}\left( q\right)}{\mathbb{E}\left[ U\left( q\right)\right]}.
\label{eq:2}
\end{equation}
Here, ${T}_{D}$, ${T}_{V}$, and ${T}_{R}$ are the mean times spent on drafting, Target verification, and correction after rejection under the measured workload. Progress is counted in terms of the output tokens ultimately committed, rather than the number of accepted Draft candidates. Increasing $q$ may advance more tokens per round, but it will also increase the work spent on candidate generation and verification, including computation for rejected candidates.

\subsection{KV Caches in Speculative Decoding}\label{sec:2.2}

Tokens produced by standard autoregressive decoding become valid history for subsequent positions. In speculative decoding, the Target KV Cache contains both committed and unverified states. During round $r$, let ${C}_{r}$ be the length of the materialized KV prefix corresponding to committed output, and let ${L}_{r}$ be the logical KV length including current candidates. Newly generated output tokens whose KV states have not yet been computed do not count toward the committed KV boundary. The state can then be written as $\mathcal K_r^{\mathrm{comm}}=[0,C_r)$ and $\mathcal K_r^{\mathrm{spec}}=[C_r,L_r)$. Here, ${\mathcal{K}}_{r}^{\mathrm{comm}}$ denotes committed history and ${\mathcal{K}}_{r}^{\mathrm{spec}}$ denotes unverified candidates. If the committed KV prefix retained at the end of the round has length ${C}_{r+1}$, the Target KV interval to reclaim is $\mathcal R_r=[C_{r+1},L_r)$. Committed KV history therefore remains valid, whereas part of the speculative frontier may be discarded. KV offloading must distinguish these two lifetimes. Consider a model with $L$ layers, ${n}_{\mathrm{kv}}$ KV heads per layer, head dimension ${d}_{h}$, and ${b}_{e}$ bytes per element. For $B$ concurrently executed requests of length $C$, the required Target KV capacity is approximately $M_{\mathrm{KV}}(B,C)\approx2BLn_{\mathrm{kv}}d_hb_eC$. The factor of 2 accounts for keys and values. Speculative decoding must additionally retain the Draft model's weights and KV Cache, as well as Target KV states for the current candidates. These states further increase memory demand at long contexts and high concurrency.

\section{Motivation}\label{sec:3}

\subsection{KV Sharing in Multi-Query Verification}\label{sec:3.1}

With KV history offloaded to CPU memory, the cost of speculative decoding depends on both the amount of KV transferred per round and the useful output that each transfer supports. For a given request, autoregressive decoding processes one new query per step, whereas a speculative Target forward pass verifies $q$ positions at once. These queries have different causal visibility within the candidate region but all access the same committed KV history. A single copy of that history can therefore supply the entire verification query group.

Sharing within a round creates an opportunity to amortize history transfers. For a fixed history length, let ${D}_{\mathrm{H}}$ be the amount of CPU-resident history read in one round and $U\left( q\right)$ the number of tokens committed. The average history transfer volume per committed token is approximately
\begin{equation}
{D}_{\mathrm{token}}\left( q\right) \approx \frac{{D}_{\mathrm{H}}}{\mathbb{E}\left[ U\left( q\right)\right]}.
\label{eq:6}
\end{equation}

The history size is largely independent of verification width, but one verification pass may advance several output tokens. Speculative verification thus reduces sequential Target calls and lets multiple useful outputs share the cost of a history transfer. Verification width affects both computation and transfer amortization. As the accepted length approaches saturation, increasing $q$ provides less transfer benefit while still adding computation. Long-context speculative serving must therefore consider history access together with the full verification round, weighing reuse against additional work in terms of cost per useful output token.

\subsection{Parallel Speculation with KV Offloading}\label{sec:3.2}
\begin{figure}[!tbp]
\centering
\includegraphics[width=0.88\linewidth]{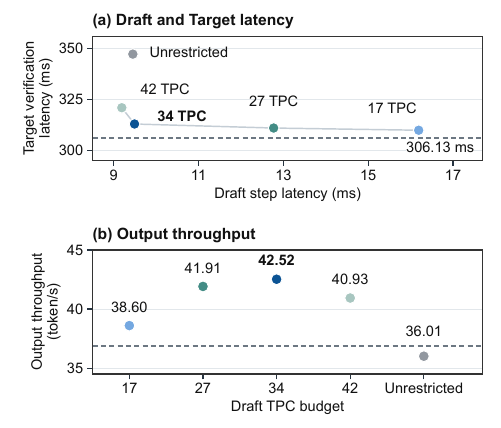}
\caption{Draft GPU resource allocation and inference performance with KV offloading. TPC~\cite{ref17} masks restrict the Draft to selected GPU compute clusters. (a)  Single-token Draft latency versus Target verification latency under different TPC allocations. (b) Corresponding output throughput. The model pair is Qwen3-32B/0.6B~\cite{ref41}, evaluated on LongBench v2. Dashed lines show the Target verification latency and end-to-end throughput of the Draft and Target serial execution. Unrestricted denotes concurrent execution on the same GPUs without a TPC limit on the Draft model.}
\label{fig:1}
\end{figure}

Parallel speculation often uses separate GPUs to overlap Draft generation with Target verification \cite{ref10,ref11,ref12}. However, memory-intensive decoding can leave spare compute capacity on the Target GPUs that these designs may not fully exploit.

KV offloading creates such an opportunity. With history stored in CPU memory, Target verification waits for the required data to arrive on the GPU. Even when transfers and computation overlap in part, the compute stream stalls if attention on the current chunk finishes before the next chunk is ready. For a stage in which transfer and computation can overlap, let ${T}_{H2D}$ denote history transfer time and ${T}_{attn}$ the attention time that can overlap it. The resulting compute wait is approximated by $G \approx \max(0,T_{\mathrm{H2D}}-T_{\mathrm{attn}})$.

Compared with keeping the entire KV Cache on the GPU, offloading introduces additional bubbles whenever H2D transfers are not hidden by computation. The copy engine is still moving data, but some compute units are idle. The models consequently have complementary resource demands during these stages. The Target is limited by data arrival, while the Draft model still has candidates to generate. Advancing the Draft model while the Target waits for history can hide part of the drafting cost within the current verification pass.

Only part of the compute capacity is available during a bubble. H2D transfers still write to GPU memory, and Target computation must resume promptly once the data arrives. Giving the Draft too many resources may interfere with transfers or delay subsequent Target kernels. Too few resources, however, may lengthen a Draft step beyond the available window. Shared-GPU execution must therefore match both the duration and resource demand of drafting to the spare capacity left by KV transfers. Only then can temporary idle capacity translate into Draft progress.

Figure~\ref{fig:1} illustrates this resource trade-off. In Fig.~\ref{fig:1}(a), increasing the Draft allocation shortens candidate generation but lengthens Target verification. With no allocation limit, the Draft step takes about as long as in some constrained configurations, yet interference with the Target is much greater. Overall throughput in Fig.~\ref{fig:1}(b) first rises and then falls; unrestricted drafting even performs below the offloading reference. Faster drafting improves overall throughput only when its benefit outweighs the interference imposed on the Target. Shared-GPU scheduling must coordinate the Draft resource allocation, start time, and work admitted per invocation to fit the available execution window.

\section{Methodology}\label{sec:4}

SpecStream jointly designs Target KV storage management, data access during verification, and intra-GPU scheduling. It comprises three components. A commit-aware KV tiering manager identifies the range eligible for migration and coordinates asynchronous copies with GPU space reclamation. A multi-query streaming verifier allows each history block to serve the entire query group and computes full attention through online softmax. It combines a transfer pipeline, Cohorts, and dynamic verification widths to improve execution efficiency. A Target-priority scheduler controls when the Draft model starts and how far it advances in parallel under the selected TPC allocation. During each round, the Target model reads the CPU-resident history and recent GPU-resident KV layer by layer, then updates the commit state after verification. The manager subsequently offloads eligible older history, while the intra-GPU scheduler advances the Draft model in permitted phases. The controller uses runtime feedback to adjust the verification width for the next round.

\subsection{Commit-Aware Rollback-Free KV Tiering}\label{sec:4.1}
\begin{figure}[!tbp]
\centering
\nativepdf[width=\linewidth]{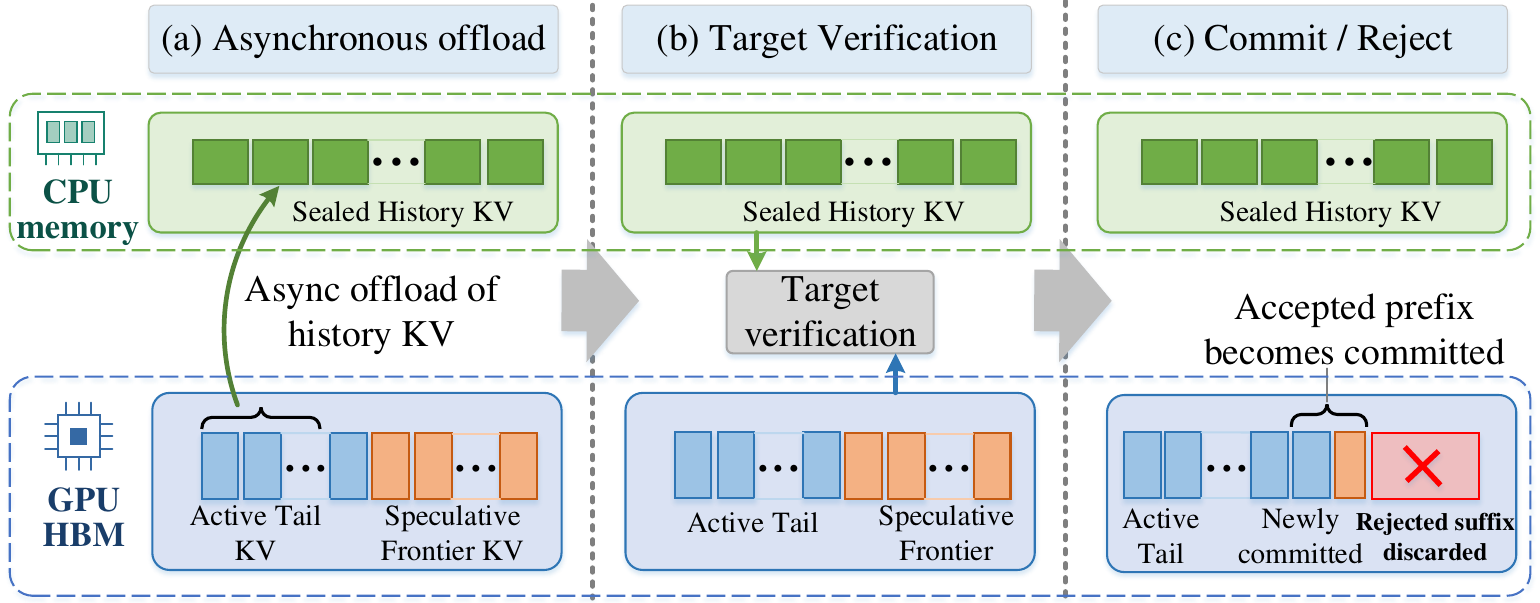}
\caption{Rollback-free KV state tiering.}
\label{fig:2}
\end{figure}

Speculative decoding assigns different lifetimes to KV states. SpecStream offloads historical KV without extending candidate rollback across devices. It aligns migration eligibility with the Target model's commit boundary: offloaded KV belongs to an irrevocably committed prefix, while KV for unverified candidates remains on the GPU. History migration and candidate rollback therefore operate on disjoint state regions, avoiding transfers of KV for candidates that may later be rejected and the need for CPU-side cleanup after rejection. Fig.~\ref{fig:2} illustrates this partitioning.

For each request, let $H$ denote the length of historical KV whose migration to the CPU has completed, ${C}_{r}$ the length of the committed KV prefix, and ${L}_{r}$ the logical length including the current candidate KV. The Target KV Cache is partitioned into the following three regions accordingly,
\begin{equation}
\begingroup
\setlength{\thickmuskip}{2mu}
\begin{array}{@{}l@{\,}l@{\,}l@{}}
\text{Sealed History}: &[0,H)   &\text{(CPU)}\\
\text{Active Tail}: &[H,C_r)    &\text{(GPU)}\\
\text{Speculative Frontier}: &[C_r,L_r) &\text{(GPU)}
\end{array}
\enspace 0 \le H \le C_r \le L_r.
\endgroup
\label{eq:8}
\end{equation}
Sealed History stores committed historical KV whose migration has been completed. Active Tail retains recently committed KV, while Speculative Frontier contains candidate KV awaiting verification. The commit boundary ${C}_{r}$ constrains the migration range, ensuring that the CPU holds only history that remains valid. Candidate acceptance, rejection, and the resulting KV reclamation are all handled on the GPU.

Committed KV need not be offloaded immediately. Offloading the KV of every newly committed token would produce many small, fragmented transfers and increase synchronization and metadata-maintenance costs. SpecStream instead retains an Active Tail with a target length of $T$ on the GPU. Newly committed KV first accumulates in this recent history, after which older contiguous KV blocks are migrated to the CPU in batches. Let the migration granularity be $g$ tokens, with $H$ already aligned to this granularity. The endpoint of the current migration is
\begin{equation}
E=\max \left\{ H,g\left\lfloor \frac{\max  \left( 0,C_r-T\right)}{g}\right\rfloor\right\}.
\label{eq:9}
\end{equation}

If $E>H$, the system migrates the interval $\left[ H,E\right)$; otherwise, it leaves the state unchanged. Migration uses asynchronous copies. The history boundary advances to $E$ only after the copy has completed and the data is confirmed to be available; the corresponding GPU space can then be reclaimed. The commit boundary ${C}_{r}$ determines which KV is eligible for migration, the recent-history retention target $T$ controls the GPU tail size, and the migration granularity $g$ controls the transfer batch size.

The verification result then determines how the speculative frontier is updated. Let ${C}_{r}^{\prime }$ be the length of the committed KV prefix retained at the end of the round. Accepted candidates move into Active Tail, and the suffix $\left[ {C}_{r}^{\prime },{L}_{r}\right)$ is released through the existing GPU KV reclamation procedure. Because later candidate rejections cannot invalidate an already committed prefix, $H\leq C_{r}\leq {C}_{r}^{\prime }$ always holds. Hence, $[C_{r}^{\prime},L_r)\cap[0,H)=\varnothing$.

The rollback range is disjoint from the CPU historical KV. Candidate rejection changes only the speculative suffix on the GPU, without requiring CPU replica cleanup or cross-device rollback. As the commit boundary advances, accepted KV gradually becomes older history and is migrated according to Eq.~\eqref{eq:9}. Using the same KV state boundaries, we next describe how CPU-resident history participates in Target verification.

\subsection{KV Streaming for Multi-Query Verification}\label{sec:4.2}
\begin{figure*}[t]
\centering
\nativepdf[width=0.95\linewidth]{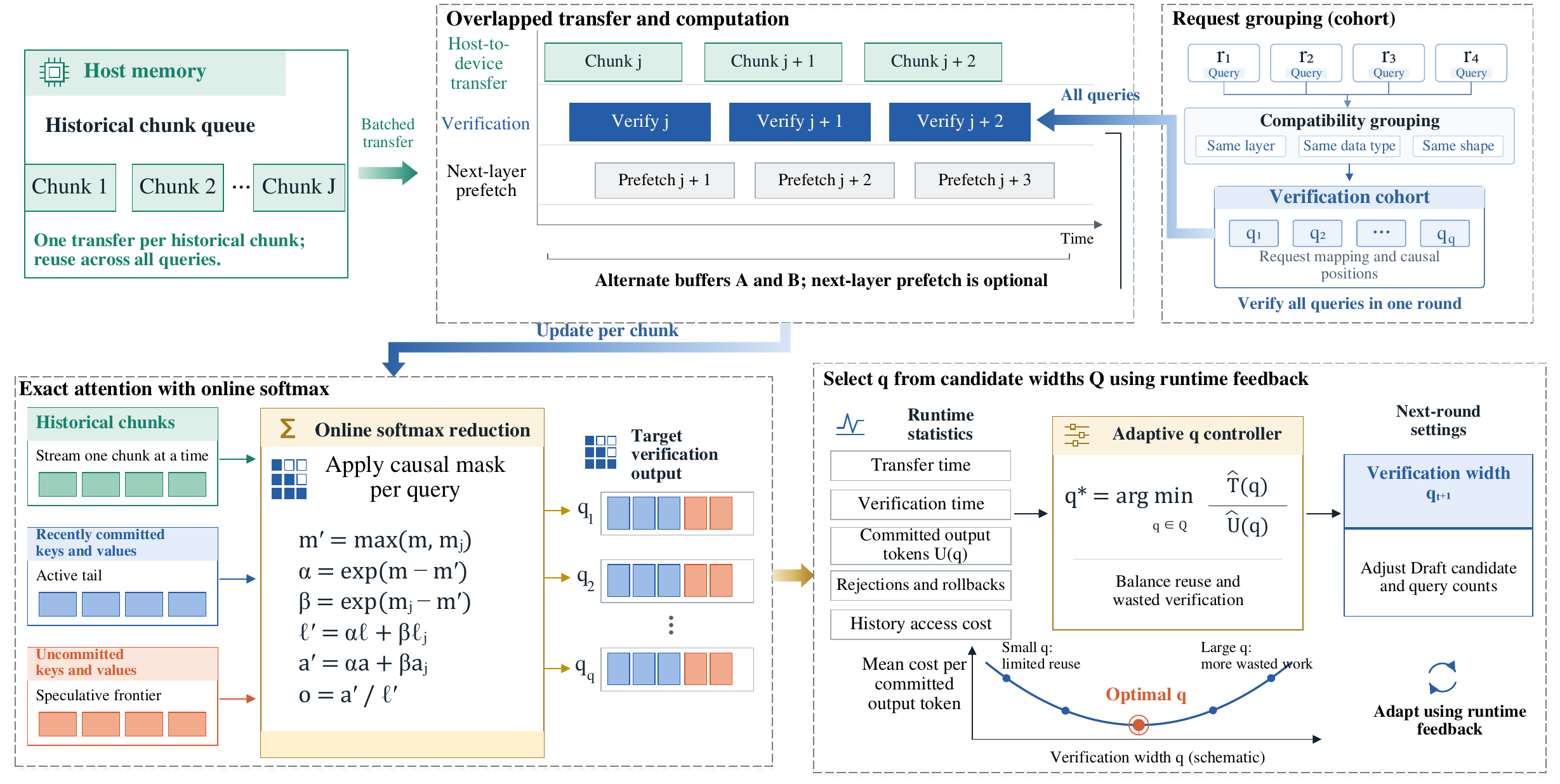}
\caption{Exact streaming execution for multi-query verification.}
\label{fig:3}
\end{figure*}

SpecStream allows Target verification to begin before the current layer's full KV history is restored. It uses the verification query group as the unit of history reuse and the KV chunk as the unit of transfer and computation. Combined with online softmax, this organization preserves full causal attention while allowing arriving history to be processed before subsequent chunk transfers complete. Double buffering and cross-layer prefetching hide part of the H2D cost, while Cohort batching reduces scheduling overhead. The system also dynamically adjusts the verification width based on the estimated time per committed token, balancing transfer amortization against additional computation. Fig.~\ref{fig:3} shows the design.

\textbf{Chunked verification shared across queries.} Although the $q$ queries in a verification round occupy different positions, they can all access the earlier Sealed History. Historical KV can therefore serve as a shared input to the entire query group, without repeated transfers for different verification positions. Visibility within the candidate region remains governed by each query's causal dependencies.

SpecStream partitions the Sealed History of the current layer into $J$ contiguous chunks along token positions. Let $Q\in {\mathbb{R}}^{q\times {d}_{h}}$ be the query matrix for the round, where ${d}_{h}$ is the head dimension. Once chunk $j$, denoted by $\left( {K}_{j},{V}_{j}\right)$, arrives on the GPU, the verifier computes its attention scores using the entire query group as $S_j=QK_j^{\mathrm T}/\sqrt{d_h}$.

A chunk needs to be transferred only once for the current layer and round to serve all $q$ verification queries. After the entire query group has finished processing it, the staging buffer can receive a subsequent chunk. Verification progresses as historical chunks arrive, without waiting for the full history of the current layer to be restored to the GPU.

The per-chunk results must also be combined into full attention. Denote the complete CPU-resident history by ${K}_{H},{V}_{H}$ and the GPU-resident Active Tail and Speculative Frontier by ${K}_{G},{V}_{G}$. The computation seeks to obtain
\begin{equation}
O=\operatorname{softmax} \left( \frac{Q{\left[ {K}_{H};{K}_{G}\right]}^{\mathrm{T}}}{\sqrt{{d}_{h}}}+M\right)\left[ {V}_{H};{V}_{G}\right].
\label{eq:12}
\end{equation}
Here, $M$ is the causal mask. For each query, all visible KV positions jointly determine the softmax normalization denominator. Normalizing each chunk independently and then summing its output would assign each chunk a different denominator, changing its relative weight in the full history. The verifier must therefore retain intermediate states that have not undergone final normalization and can be merged across chunks.

SpecStream uses online softmax to maintain normalization state chunk by chunk \cite{ref13,ref14}. For each query and attention head, it stores three accumulators $\left( m,\ell ,a\right)$. Here, $m$ is the maximum attention score over the KV positions processed so far, $\ell $ is the sum of exponentials relative to that maximum, and $a$ is the weighted sum of values using the same exponential weights. The latter two accumulate the denominator and numerator of the final output, respectively, without performing the division during intermediate computation. The initial state is $\left( -\infty ,0,0\right)$.

For each newly arriving chunk $j$, let ${S}_{j}$ denote its attention scores and ${V}_{j}$ its values. The following updates are performed separately for each query and attention head, with $k$ ranging over the KV positions in the current chunk that are visible to that query. The verifier first computes the local state within the chunk:
\begin{equation}
\textstyle m_j=\max(S_j),\ \ell_j=\sum_k e^{S_{j,k}-m_j},\ a_j=\sum_k e^{S_{j,k}-m_j}V_{j,k}.
\label{eq:13}
\end{equation}
Here, ${m}_{j}$ is the maximum score in the current chunk, while ${\ell }_{j}$ and ${a}_{j}$ are its local sum of exponentials and weighted sum of values, both relative to ${m}_{j}$. The accumulated state and the current chunk may use different maxima, so they must be rescaled to a common reference before merging. Let
\begin{equation}
m'=\max(m,m_j),\quad\alpha=e^{m-m'},\quad\beta=e^{m_j-m'}.
\label{eq:14}
\end{equation}
The accumulated denominator and weighted sum of values are then updated as $\ell^{\prime}=\alpha\ell+\beta\ell_j$ and $a^{\prime}=\alpha a+\beta a_j$.

The scaling factors $\alpha $ and $\beta $ align the existing state and the current chunk's local state to the common maximum ${m}^{\prime }$. Because the numerator and denominator use the same scaling factors, merging preserves the relative weights of previously processed KV in the full softmax. After processing each chunk, the system sets $\left( m,\ell ,a\right)\leftarrow \left( {m}^{\prime },{\ell }^{\prime },{a}^{\prime }\right)$ and proceeds to the next chunk.

After processing all CPU history chunks, the verifier continues with Active Tail and Speculative Frontier on the GPU. The earlier CPU-resident history is visible to the entire query group. The GPU-resident region, however, requires a causal mask based on each query's absolute position to exclude subsequent candidates that it cannot see. These results update the accumulated state in the same manner. Once all causally visible KV has been processed, the attention output is $O=a/\ell$.

Each chunk's result has been rescaled to a common maximum during merging. The final accumulators therefore satisfy
\begin{equation}
\frac{a}{\ell }=\frac{\sum_{t\in \mathcal{V}} {e}^{{s}_{t}-m}{V}_{t}}{\sum_{t\in \mathcal{V}} {e}^{{s}_{t}-m}}.
\label{eq:17}
\end{equation}
Here, $\mathcal{V}$ denotes all KV positions causally visible to the current query. This ratio uses a numerator and denominator jointly formed over all visible positions, making the chunked reduction equivalent to computing the full softmax at once in real arithmetic \cite{ref13}. SpecStream changes only the order in which KV arrives on the GPU and participates in the reduction. It preserves the definition of full-history attention.

\textbf{Pipelined KV transfers and prefetching.} Chunked reads shorten the wait before computation can begin. Yet if every chunk is processed serially, with each transfer followed by computation and the next transfer starting only afterward, the waiting time still accumulates across chunks. SpecStream uses two GPU staging buffers and separate transfer and compute streams to overlap H2D transfers with attention computation. While the GPU processes chunk $j$ in buffer A, buffer B receives chunk $j+1$. Once computation on the current chunk finishes, A receives subsequent data. Adjacent chunks are grouped into transfer batches to amortize copy-submission, synchronization, and kernel-launch overheads. Once the current layer releases a reusable buffer, the system prefetches historical KV for the next layer, overlapping the transfer with the current layer's remaining attention, output projection, and feed-forward computation to reduce waits in the next layer.


\textbf{Cohort batching under high request concurrency.} When multiple requests enter verification together, processing historical chunks separately for each request creates many fine-grained tasks. Each request requires data preparation, transfer submission, and attention kernel launches at every layer. Even if an individual task overlaps transfer with computation, preparation and synchronization costs still accumulate with the request count, and small compute tasks may underutilize the GPU. SpecStream constructs Cohorts within the current scheduled batch, grouping tasks that can be processed by the same batched kernel. Compatibility requires matching model layers, verification widths, data types, numbers of KV heads, head dimensions, and chunk sizes. The system prepares historical chunks and queries by group, then executes each request's computation through a batched attention kernel. Requests retain independent KV indices, absolute positions, and online softmax states, and each result updates its own accumulators. This increases the parallelism of each kernel invocation while amortizing data preparation, synchronization, and kernel-launch costs.

\textbf{I/O-aware dynamic verification widths.} Sharing historical KV across queries makes the verification width $q$ determine both the candidate count and the granularity of history reuse. The benefit of reuse also depends on the number of tokens actually committed. Following the transfer amortization relationship in Eq.~\eqref{eq:6}, SpecStream dynamically selects $q$ using candidate acceptance and execution costs.

With a long KV history and a small verification width, each read is amortized over only a few output tokens. A moderate increase in $q$ allows the same history to serve more potentially useful outputs; when transfers are the bottleneck, the additional attention computation may also be hidden by transfer time. Once the accepted length approaches saturation, this benefit diminishes even as candidate generation and verification work continue to grow. The controller therefore selects $q$ by the estimated time per useful output token. It uses recent history-transfer, verification, Draft-generation, and rejection-correction costs, together with the number of tokens actually committed, to estimate the round cost at different widths and selects $q^{\ast}=\operatorname*{arg\,min}_{q\in\mathcal Q}(\widehat T_{\mathrm{round}}(q)/\widehat U(q))$. Here, $\mathcal{Q}$ is the set of candidate verification widths, and ${\widehat{T}}_{\mathrm{r}\mathrm{o}\mathrm{u}\mathrm{n}\mathrm{d}}\left( q\right)$ and $\widehat{U}\left( q\right)$ denote the estimated round time and number of committed tokens, respectively. 

The controller records Draft execution time and useful progress for each width, prioritizing existing observations. When records are insufficient, it estimates costs from available measurements and collects information for each candidate width through a limited initial sampling phase. Verification costs are estimated from the historical data volume, transfer rate, and chunk computation time, without counting overlapped time twice. The system updates its statistics at the end of each round and changes the width for the next round only when the estimated benefit exceeds the switching threshold. Thus, $q$ adapts to history-access costs, batching conditions, and candidate quality while limiting frequent switching.


\subsection{Target-Priority Intra-GPU Co-Execution}\label{sec:4.3}
\begin{figure*}[t]
\centering
\nativepdf[width=0.95\linewidth]{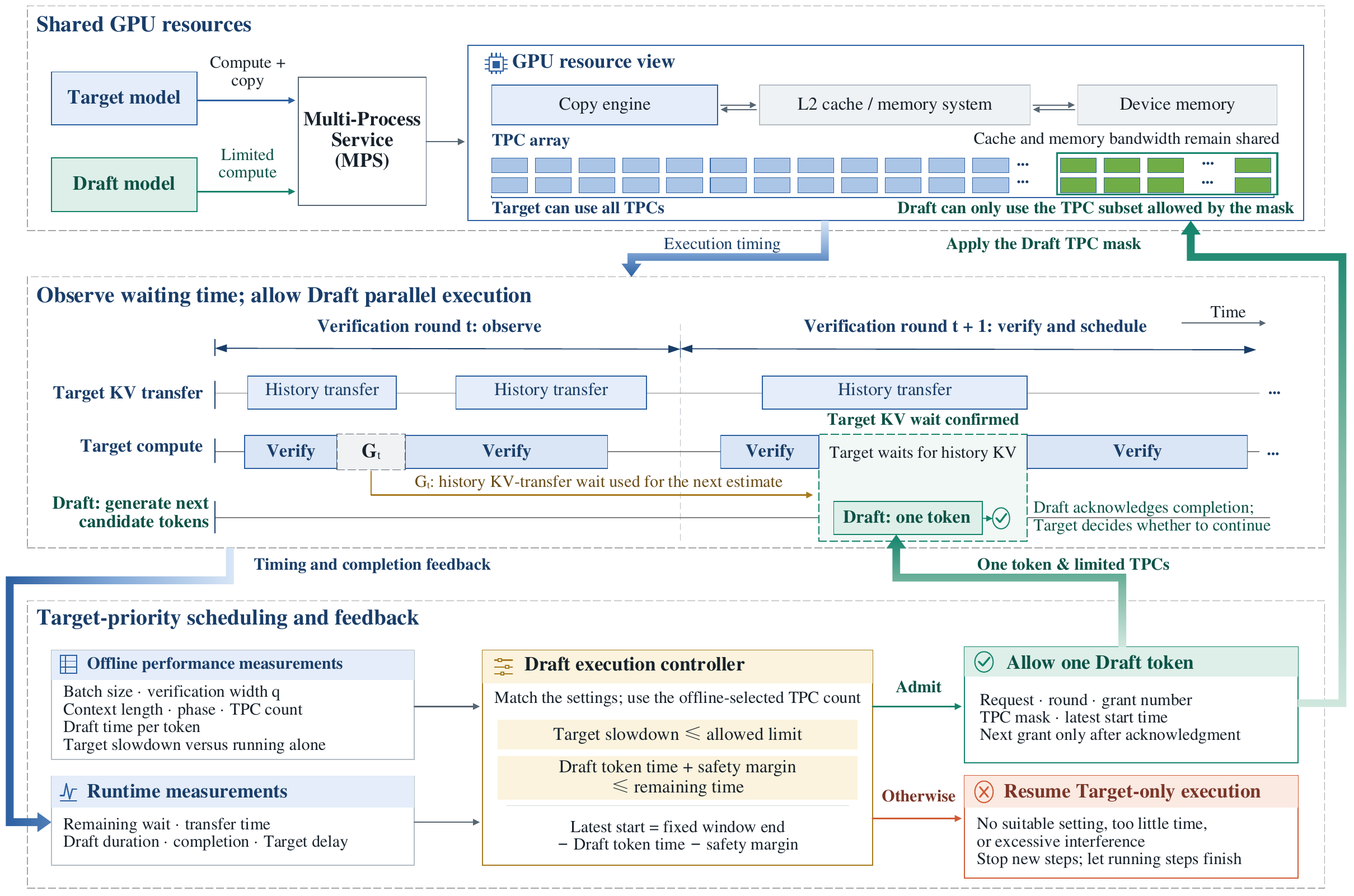}
\caption{Target-Priority Parallel Speculation on Shared GPUs.}
\label{fig:4}
\end{figure*}

SpecStream turns the compute bubbles caused by Target KV-transfer waits into candidate-generation progress on the same GPU. This enables parallel speculation without provisioning additional compute resources for the Draft model. The scheduler uses the Target model's available time and latency budget to govern Draft execution, rather than simply pursuing the fastest possible drafting. An offline-selected TPC allocation limits Draft resource use, while runtime admission determines when the allocated resources can be used within the remaining window. Fig.~\ref{fig:4} shows the design.

\textbf{Draft compute allocation.} SpecStream uses NVIDIA Multi-Process Service (MPS) \cite{ref15} to enable concurrent execution of the Target and Draft models in separate processes. MPS allows kernel execution and data copying from different CUDA processes to overlap on the same GPU, providing an environment for the two models to execute concurrently. For resource control, SpecStream limits the Draft model's compute resources at TPC granularity. A TPC is a hardware cluster containing Streaming Multiprocessors (SMs) \cite{ref16}. The system applies a TPC mask in the separate Draft process so that its compute kernels run only on the designated set of TPCs \cite{ref17}, while preserving the Target model's original access to compute resources. TPC restrictions do not isolate the L2 cache or memory bandwidth, so resource contention between the models must still be controlled at runtime.

At deployment, SpecStream first measures the execution performance of both models under different TPC allocations. With fewer TPCs, the Draft model takes longer to generate one token; with more, it may prolong Target verification. The system defines a runtime configuration $x$ by the model pair, batch size, context length, verification width, and execution phase. It records the Draft model's single-step time ${t}_{D}\left( x,n\right)$ with $n$ TPCs and computes the increase in Target verification latency relative to execution alone as $\Delta_{\mathrm T}(x,n)=[T_{\mathrm T}^{\mathrm{co}}(x,n)-T_{\mathrm T}^{\mathrm{alone}}(x)]/T_{\mathrm T}^{\mathrm{alone}}(x)$. Here, ${T}_{\mathrm{T}}^{\mathrm{co}}\left( x,n\right)$ and ${T}_{\mathrm{T}}^{\mathrm{alone}}\left( x\right)$ are the Target verification latencies during co-execution with the Draft model and during standalone execution, respectively. Both include the wait for historical KV. The Draft model's single-step time is also measured during co-execution to capture the effect of contention on candidate generation. The offline sweep additionally compares end-to-end throughput across allocations and selects a default TPC allocation ${n}_{0}$. At runtime, the system determines whether the Draft model may start under this allocation.

\textbf{Target-priority scheduling.} At runtime, SpecStream uses the transfer wait, verification-phase durations, and Draft single-step time observed in the preceding round to update the corresponding time estimates for the next round. It then estimates the remaining time available to the Draft model, $\widehat{S}$, from current progress. Starting from the offline records, the system incorporates runtime feedback to update the estimated Draft single-step time ${\widehat{t}}_{\mathrm{D}}\left( x,{n}_{0}\right)$ and Target latency increase ${\widehat{\Delta }}_{\mathrm{T}}\left( x,{n}_{0}\right)$ under the selected allocation. Let ${\varepsilon }_{\mathrm{T}}$ be the acceptable increase in Target latency and $\delta $ a time margin reserved for execution variability. Starting a Draft step under allocation ${n}_{0}$ must satisfy conditions: $\widehat\Delta_{\mathrm T}(x,n_0)\le\varepsilon_{\mathrm T}$ and $\widehat t_{\mathrm D}(x,n_0)+\delta\le\widehat S$. The first condition bounds the estimated increase in Target latency. The second requires the estimated duration of one Draft step, including the reserved margin, to fit within the remaining window. The scheduler submits a new concurrent step only when both conditions hold; otherwise, it defers execution.

The available window changes as the Target model advances. The scheduler therefore admits only one full Draft forward pass for a single token at a time. Longer windows can accommodate several successive steps; shorter ones permit fewer steps or none. Single-token admission bounds the in-flight work that cannot be revoked immediately, reducing the risk that Draft execution outlasts the waiting window and delays Target computation.

Each scheduling instruction carries the request and speculative-round identifiers, an execution sequence number, the TPC allocation, and the latest start time. If the instruction is issued at $t_0$ with an estimated remaining time of $\widehat S$, its latest start time is $t_{\mathrm{latest}}=t_0+\widehat S-\widehat t_{\mathrm D}(x,n_0)-\delta$. Before starting, the Draft model checks that the instruction still matches the current request and speculative round and that the latest start time has not passed. Expired or mismatched instructions are not executed. After completing the GPU computation for one token, the Draft model returns an acknowledgment (ACK), with the execution sequence number linking the completion result to the corresponding instruction. Only after receiving the ACK does the Target model decide whether to schedule another step based on the remaining time, keeping at most one unfinished Draft step per request.

Execution results are then used to update the estimated Draft single-step time and Target latency. When the dynamic verification-width controller in Section~\ref{sec:4.2} changes $q$, or when the batch size or context length changes, the system reads the measurement records for the default allocation under the new runtime configuration. If the observed increase in Target latency exceeds the budget, the scheduler stops submitting new concurrent steps and restores exclusive Target execution once the already launched Draft computation finishes. If the Target model is already waiting for candidate tokens required by the current round, the request continues through the original Draft-generation procedure. Whether parallel execution resumes is determined by the Target model's current state.

\section{Experimental Results}\label{sec:5}


\subsection{Experimental Setup}\label{sec:5.1}

\textbf{Models and deployment.} We implement SpecStream on SGLang 0.5.17 \cite{ref18} and evaluate two Target/Draft model pairs, Qwen3-32B/0.6B \cite{ref41} and InternLM2.5-20B-Chat/1.8B-Chat \cite{ref42}. The main experiments use two NVIDIA A800 80GB PCIe Gen4 GPUs and 240GB of host memory. SpecStream's Target and Draft models share these GPUs, both using tensor parallelism (TP) with degree 2, and use separate process groups and MPS to support concurrent execution. The separate-GPU parallel baseline in Section~\ref{sec:5.3.3} uses three A800 GPUs. Two run the Target with TP=2, and the third runs a dedicated Draft model with TP=1.

\textbf{Workloads.} GSM8K \cite{ref19} evaluates reasoning with short inputs, while LongBench v2 \cite{ref20} and MRCR \cite{ref21} cover long-document understanding and retrieval from long conversations, respectively. The component ablations use token-ID inputs generated with a fixed random seed to control input length and the number of concurrent requests.

\textbf{Baselines.} We compare SpecStream with autoregressive decoding (AR), SGLang speculative decoding (SGLang SD) \cite{ref18}, SGLang SD+Offload (SD+Offload), and TriForce \cite{ref32}. SD+Offload uses chunked KV transfers, double buffering, and cross-layer prefetching, but starts attention only after the layer's full KV history is restored to the GPU. TriForce combines a Draft model with a bounded KV window, retrieval-based intermediate speculation, and verification against the full history. Because its public implementation targets Llama models, we adapt the original method to Qwen3 and InternLM2.5. Related systems emphasize different designs: VeriCache \cite{ref33} uses compressed-KV self-speculation, SpecOffload \cite{ref34} targets weight-offloaded inference, and Dovetail \cite{ref35} performs Target verification on the CPU. We select TriForce as a representative offloading baseline for our setting. SGLang SD and SD+Offload use a fixed verification width of $q=4$, whereas SpecStream dynamically selects $q$ from \{2, 4, 6, 8\}. Unless otherwise specified, SpecStream retains a 512-token Active Tail and allocates no additional GPU cache for historical Target KV. For parallel speculation, we also compare against SPECTRE+Offload, which uses SPECTRE's \cite{ref11} parallel execution mechanism across separate GPUs, together with our history offloading and streaming verification path.

\textbf{Metrics.} Our primary performance metric is end-to-end output throughput in tokens/s. Task quality is measured using numerical-answer accuracy on GSM8K, accuracy on LongBench v2, and token F1 on MRCR.

\subsection{End-to-End Performance}\label{sec:5.2}

\subsubsection{Inference Accuracy}\label{sec:5.2.1}

In Table~\ref{tab:1}, the largest score difference between SpecStream and SGLang SD is 0.1988 percentage points across the two model pairs and three tasks. SpecStream thus maintains output quality close to that of SGLang SD on the evaluated tasks. As analyzed in Section~\ref{sec:4.2} for full causal attention, SpecStream's streaming verification changes only the order of KV access and reduction. The different floating-point accumulation order introduced by chunked reduction may cause small numerical perturbations, resulting in differences in accuracy.

\begin{table}[!tbp]\centering
\caption{Task-quality scores of SpecStream and SGLang SD.}
\label{tab:1}
\small\setlength{\tabcolsep}{2pt}
\resizebox{\linewidth}{!}{%
\begin{tabular}{@{}llrrr@{}}
\toprule
\textbf{Model} & \textbf{Dataset} & \shortstack{\textbf{SGLang SD} \textbf{(\%)}} & \shortstack{\textbf{SpecStream} \textbf{(\%)}} & \shortstack{\textbf{Difference} \textbf{(pp)}} \\
\midrule
\multirow{3}{*}{Qwen3} & GSM8K & 95.2237 & 95.1479 & $-0.0758$ \\
 & LongBench v2 & 45.8015 & 45.8015 & $0.0000$ \\
 & MRCR & 55.4946 & 55.6571 & $+0.1625$ \\
\addlinespace[2pt]
\multirow{3}{*}{InternLM2.5} & GSM8K & 84.9886 & 84.9886 & $0.0000$ \\
 & LongBench v2 & 40.3509 & 40.1521 & $-0.1988$ \\
 & MRCR & 52.3585 & 52.2939 & $-0.0646$ \\
\bottomrule
\end{tabular}}
\end{table}

\subsubsection{Performance Evaluation}\label{sec:5.2.2}
\begin{figure}[t]
\centering
\includegraphics[width=0.92\linewidth]{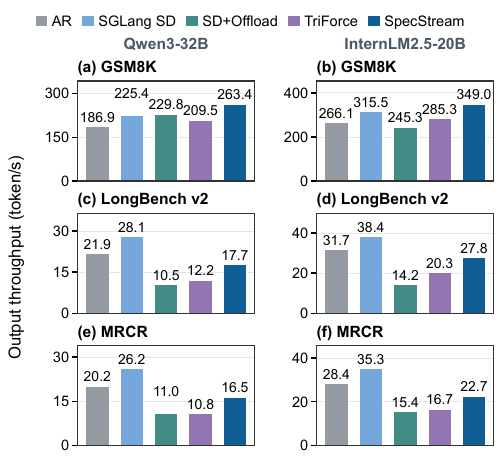}
\caption{End-to-end output throughput on different datasets. The number of concurrent requests is 8.}
\label{fig:5}
\end{figure}

Fig.~\ref{fig:5} compares end-to-end throughput across methods. SD+Offload and SpecStream retain only the most recent 512 tokens (Active Tail) of Target KV on the GPUs. TriForce maintains a GPU retrieval cache for intermediate speculation and stores Target KV according to the offloading configuration used in this experiment. Across the three datasets, SpecStream achieves average throughput speedups of 1.41\ensuremath{\times} and 1.32\ensuremath{\times} over TriForce on Qwen3 and InternLM2.5, respectively. The corresponding speedups over SD+Offload are 1.44\ensuremath{\times} and 1.62\ensuremath{\times}. Streaming verification of SpecStream is a key source of these gains. Each incoming history KV chunk immediately serves the entire query group in the round, reducing data stalls during verification. TriForce, by contrast, reduces the frequency of full-history verification through hierarchical speculation, but incurs additional intermediate computation and retrieval-cache maintenance costs. SpecStream further improves throughput by dynamically adjusting the verification width and using spare compute capacity during transfers to advance the Draft in parallel. On the long-input LongBench v2 and MRCR workloads, however, SpecStream's throughput remains below that of AR and SGLang SD with fully GPU-resident KV. Streaming verification can hide part of the transfer wait, but historical KV must still traverse PCIe in every verification round. The unidirectional bandwidth of PCIe Gen4$\times$16 is approximately 31.5~GB/s, far below the A800's HBM bandwidth of 1935~GB/s. Offloading the long KV history beyond the Active Tail to CPU memory reduces GPU memory usage but increases the cost of historical KV access during Target verification. The shorter inputs in GSM8K make historical KV access less costly. Here, SpecStream improves throughput over SGLang SD by 16.9\% and 10.6\% for the two model pairs, respectively. Its dynamic verification width increases the mean accepted length by approximately 16\% and 28\% for the Qwen3 and InternLM2.5 model pairs, respectively, allowing each verification round to advance more output tokens.

\begin{figure*}[t]
\centering
\includegraphics[width=0.82\linewidth]{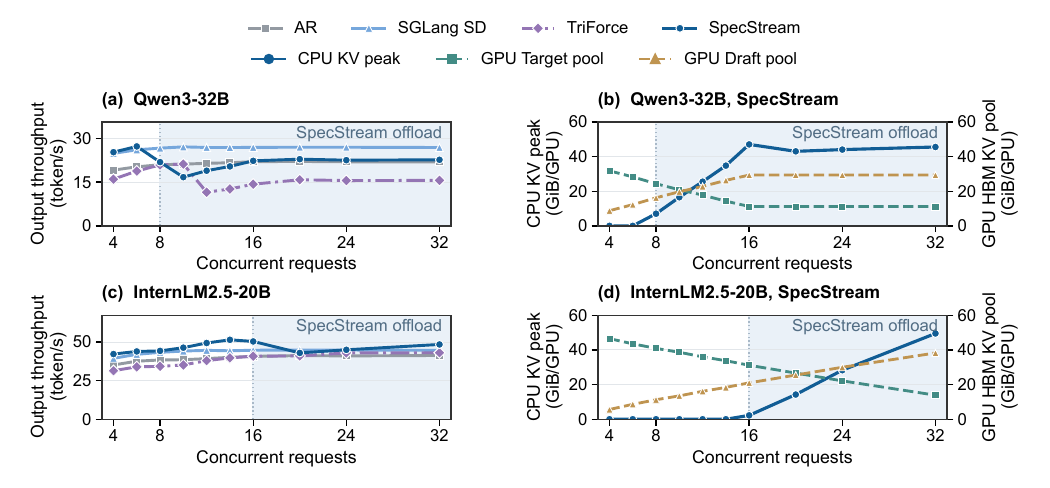}
\caption{Performance under increasing concurrency and KV storage distribution with GPU-first residency. The left column compares end-to-end throughput. The right column shows SpecStream's peak CPU KV history usage and GPU KV pool capacity, both reported per GPU. Shading marks the concurrency range in which SpecStream offloads KV states.}
\label{fig:6}
\end{figure*}

To examine how history offloading affects serving performance under limited GPU memory, we use a GPU-first residency policy on LongBench v2 and gradually increase the number of concurrent requests. Offloading begins only when historical KV states exceed the available capacity. After accounting for model weights and other buffers, SpecStream's combined GPU KV pool is fixed at 40.50 GiB per GPU for Qwen3 and 52.15 GiB per GPU for InternLM2.5. TriForce applies the same GPU-first residency policy to its full Target KV Cache while retaining the retrieval cache required for intermediate speculation. Fig.~\ref{fig:6} compares throughput across methods and reports SpecStream's peak CPU KV history usage and GPU KV pool capacity.

Unlike AR and SGLang SD, which keep the full KV Cache on the GPUs, SpecStream and TriForce alleviate GPU memory limits by offloading historical KV states. As concurrency increases, GPU memory limits force additional requests to queue, causing AR and SGLang SD throughput to saturate. SpecStream moves the Target's KV history to CPU memory, preserving GPU capacity for active batches. With 32 concurrent requests, its peak CPU KV history usage reaches 45.50 and 49.50 GiB per GPU for the two model pairs. This capacity benefit comes with a transfer cost. SpecStream begins offloading at 8 concurrent requests on Qwen3 and 16 on InternLM2.5. Throughput drops and then recovers in some of the subsequent concurrency configurations. At 32 concurrent requests, Qwen3 throughput is slightly higher than AR but lower than SGLang SD, whereas InternLM2.5 exceeds them by 17.3\% and 8.6\%, respectively. Cumulative H2D traffic for verification at this point is 7.79 and 3.29 TiB per GPU for the two model pairs. Although their peak CPU KV history usage is similar, Qwen3 transfers substantially more data overall, consistent with its weaker throughput recovery.

SpecStream achieves average throughput speedups of 1.40\ensuremath{\times} and 1.23\ensuremath{\times} over TriForce on Qwen3 and InternLM2.5, respectively. TriForce's bounded-window Draft model reduces the small model's resident KV footprint. Its retrieval-based intermediate speculation reduces the number of full verification passes, while introducing retrieval-cache maintenance and intermediate computation. SpecStream focuses on reducing data stalls and execution overhead in full Target verification, achieving higher average throughput in these tests. It does, however, retain the complete Draft KV Cache on the GPUs. As the number of concurrent requests grows, this cache occupies an increasing share of GPU memory and becomes a non-negligible cost, even though its initial footprint is small. It therefore also limits the number of requests that SpecStream can process simultaneously on the GPUs.

\subsection{Ablation Study}\label{sec:5.3}

\subsubsection{Streaming Transfers, Prefetching, and Cohort}\label{sec:5.3.1}
\begin{figure}[t]
\centering
\includegraphics[width=\linewidth]{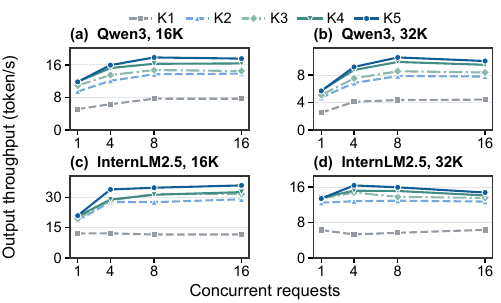}
\caption{Throughput changes in the ablation of SpecStream's streaming verification.}
\label{fig:7}
\end{figure}

To separate the contributions of individual mechanisms, we first disable concurrent execution on shared GPUs and progressively enable the optimizations in Table~\ref{tab:2} within the same SpecStream implementation. K1 through K5 use identical KV budgets and inputs, with K5 representing the complete streaming verification scheme. We test 16K and 32K inputs with 1, 4, 8, and 16 concurrent requests, fixing each request's output at 256 tokens. Fig.~\ref{fig:7} presents the results.

\begin{table}[t]\centering
\caption{Cumulative ablation configurations for SpecStream.}
\label{tab:2}
\small\setlength{\tabcolsep}{4pt}
\resizebox{\linewidth}{!}{%
\begin{tabular}{@{}llr@{}}
\toprule
\textbf{Configuration} & \textbf{Change from the preceding configuration} & $q$ \\
\midrule
K1 & Verify after restoring the full history & 4 \\
K2 & Stream-and-reduce arriving KV chunks & 4 \\
K3 & Cross-chunk and cross-layer prefetching & 4 \\
K4 & Dynamic verification width & 2, 4, 6, 8 \\
K5 & Merge compatible requests into a Cohort & 2, 4, 6, 8 \\
\bottomrule\end{tabular}}\end{table}

In Fig.~\ref{fig:7}, streaming access for the full query group yields the largest throughput gain. With 16K inputs, K2 improves throughput over K1 by an average of 83.2\% on Qwen3 and 117.1\% on InternLM2.5. Both use $q=4$; they differ in how they organize access to the KV history. K2 uses each arriving KV chunk directly to compute attention for the entire query group, reducing fine-grained copy and submission overhead and starting verification before the full history has been restored. K3 additionally prefetches subsequent chunks and the next layer's history. It improves throughput over K2 by 4.5\% to 15.6\% and 3.0\% to 15.3\%, respectively, showing that chunking alone leaves data stalls that prefetching can hide. K4 dynamically adjusts the verification width, changing throughput by 8.5\% to 16.0\% on Qwen3 and \ensuremath{-}1.2\% to 9.6\% on InternLM2.5. Its benefit depends on the relationship between additional computation and useful progress, which Section~\ref{sec:5.3.2} examines further. Cohort primarily benefits from merging work across requests. With 4 to 16 concurrent requests, K5 improves throughput over K4 by 4.7\% to 9.6\% and 4.6\% to 17.3\%, respectively. It brings almost no gain with a single request, where opportunities to share batched submissions and kernel execution across compatible requests are limited. Overall, K5 achieves 2.23\ensuremath{\times} to 2.51\ensuremath{\times} and 1.73\ensuremath{\times} to 3.09\ensuremath{\times} the throughput of K1 on the two model pairs.

\subsubsection{Dynamic Verification Width}\label{sec:5.3.2}
\begin{figure*}[t]
\centering
\includegraphics[width=0.80\linewidth]{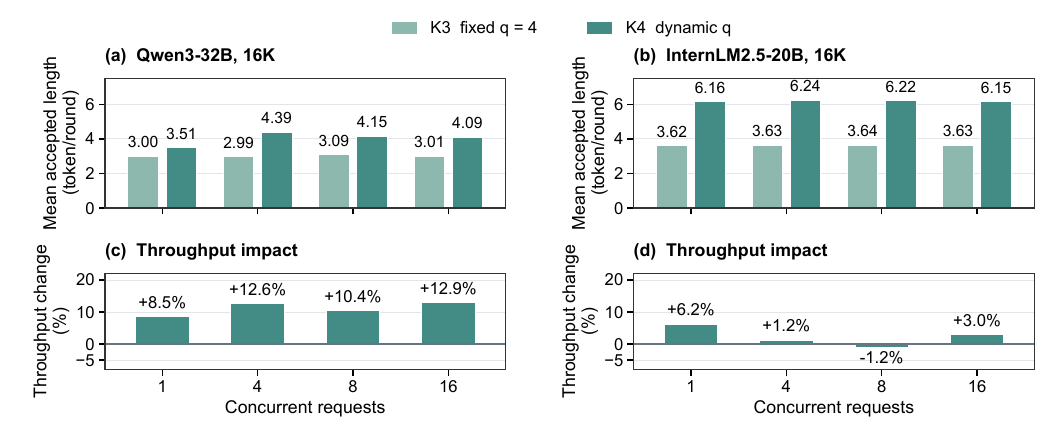}
\caption{Effect of dynamic verification width on mean accepted length and throughput. The top row compares mean accepted lengths for K3 and K4; the bottom row reports K4's throughput change relative to K3.}
\label{fig:8}
\end{figure*}

Fig.~\ref{fig:8} shows the changes in mean accepted length and throughput from K3 to K4. A longer mean accepted length does not guarantee higher throughput. On Qwen3 with 16K inputs and four requests, the mean accepted length increases by 46.8\%, but throughput improves by only 12.6\%. On InternLM2.5 with eight requests, the mean accepted length rises from 3.64 to 6.22, yet throughput falls by 1.2\%. A larger verification width $q$ increases computation for both drafting and verification, and rollback after rejection may introduce more wasted work. The extent to which historical KV reads can be amortized also varies with context length, batch size, and acceptance behavior. Mean accepted length alone therefore cannot determine $q$. SpecStream compares the estimated round time and useful progress at different values of $q$, seeking a width that balances the two to reduce the cost per output token.

\subsubsection{Target-Priority Scheduling on Shared GPUs}\label{sec:5.3.3}
\begin{figure}[!htbp]
\centering
\includegraphics[width=0.75\linewidth]{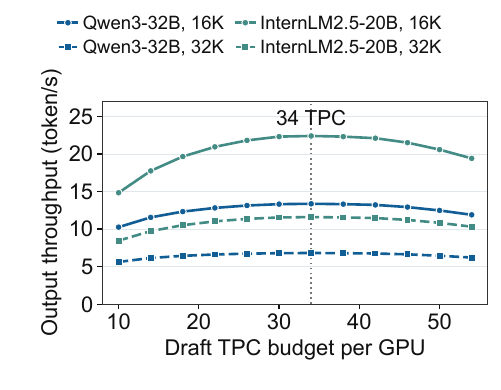}
\caption{Effect of the Draft TPC allocation on throughput.}
\label{fig:9}
\end{figure}
\begin{figure}[!tbp]
\centering
\includegraphics[width=0.85\linewidth]{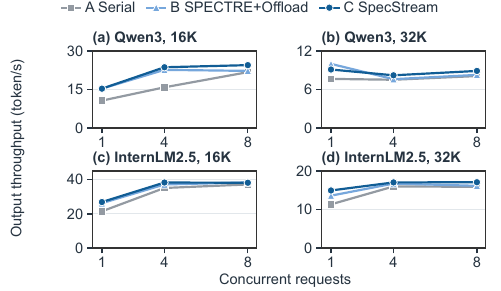}
\caption{End-to-end output throughput of sequential execution, SPECTRE+Offload, and SpecStream.}
\label{fig:10}
\end{figure}

We first determine the Draft model's default TPC allocation through an offline sweep, then evaluate the overall benefit of scheduling concurrent execution on shared GPUs. For both model pairs, we measure end-to-end throughput as the Draft allocation increases from 10 to 54 TPCs per GPU. Fig.~\ref{fig:9} shows a similar tradeoff across input lengths for both pairs. Increasing the allocation improves throughput in the low-TPC range, followed by a relatively flat region of high throughput around 30 to 38 TPCs. The highest value in this sweep occurs at 34 TPCs. This agrees with the latency trend in Fig.~\ref{fig:1}. With a small allocation, a Draft step takes longer, limiting candidate generation. Once the allocation is sufficient, further resources provide diminishing returns and may increase interference with the Target. We therefore use a default allocation of 34 TPCs per GPU in the concurrent execution experiments. At runtime, the scheduler controls when drafting starts and how much work it performs according to the Target's state.

Fig.~\ref{fig:10} compares three execution modes. A executes the Draft and Target sequentially, while C uses Target-priority concurrent execution on the same GPUs. Both use two GPUs with TP=2 for both models. B is SPECTRE+Offload, which uses three GPUs with TP=2 for the Target and TP=1 for a dedicated Draft model. Across paired configurations with the same batch size, C improves throughput over A by 9.9\% to 43.6\% and 2.7\% to 31.7\% for the two model pairs. The relative gain is largest with a single request. With less parallel work within the batch, advancing the Draft earlier helps shorten the wait between drafting and verification. As the number of concurrent requests increases, Target batching becomes more effective, and the scheduling benefit generally narrows. This indicates that the amount of concurrent Draft work needs to adapt to the Target's load.

Across the two model pairs, SpecStream achieves an average throughput speedup of 1.036\ensuremath{\times} over SPECTRE+Offload. Because SPECTRE uses one additional GPU, normalizing by GPU count gives SpecStream an average improvement of 55.4\% in output throughput per GPU. Both schemes overlap drafting with Target verification, but they use different resources. SpecStream reuses spare compute capacity while the Target waits for historical KV. Execution on separate GPUs provides dedicated compute resources for the Draft model, while requiring candidates and verification results to be exchanged across devices. Once candidates can be supplied on time, faster drafting need not shorten a round dominated by historical KV reads and Target verification. Scheduling concurrent execution on shared GPUs can therefore sustain comparable throughput with fewer devices and improve average output throughput per GPU in this experimental configuration.

\subsection{Discussion}\label{sec:5.4}

The results show that SpecStream alleviates capacity limits by tiering historical KV states and improves throughput under offloading through multi-query streaming verification and scheduling on shared GPUs. Streaming verification promptly uses each arriving history chunk for all queries in the round, reducing exposed data stalls. Concurrent scheduling advances the Draft with spare compute capacity during transfers, allowing two GPUs to achieve throughput close to three-GPU SPECTRE+Offload. These gains come from coordinating KV access with model execution. However, Fig.~\ref{fig:6} also shows that expanding the KV pool for the Draft reduces the GPU residency budget for the Target's KV. The current implementation retains the complete Draft KV on GPU, because consecutive Draft steps lack the opportunity for multi-query reuse within a Target verification round. Directly applying the same offloading path could require repeatedly fetching historical KV across consecutive steps, increasing the wait for candidate generation. Expanding the active batch further therefore calls for Draft KV management that uses less GPU memory.

\section{Related Work}\label{sec:6}

\textbf{Speculative decoding and verification optimization.} Speculative decoding reduces sequential Target invocations through inexpensive drafting and batched verification \cite{ref03,ref04}. Medusa \cite{ref05}, the EAGLE family \cite{ref06,ref22,ref23}, SpecInfer \cite{ref07}, and Sequoia \cite{ref24} improve useful output per round through candidate prediction and tree-based verification. MagicDec \cite{ref25} and QuantSpec \cite{ref26} reduce long-context drafting overhead using sparse KV states or quantization. DeFT \cite{ref27} groups queries and KV states around shared prefixes to reduce redundant KV reads within the GPU. 


\textbf{KV management and offloading.} Paged allocation \cite{ref01}, prefix sharing \cite{ref18}, and context caching \cite{ref08,ref28,ref29} reduce wasted KV storage and repeated prefill computation. FlexGen \cite{ref02} jointly plans the tiered placement of weights, activations, and KV states. HeadInfer \cite{ref30} offloads KV states at attention-head granularity, whereas InfiniGen \cite{ref09} and SpeCache \cite{ref31} selectively fetch important KV states to reduce data movement. Recent work also combines offloading with speculative decoding. TriForce \cite{ref32} builds hierarchical speculation around retrieval-based drafting and amortizes the cost of fetching the full history through multi-token verification. VeriCache \cite{ref33} drafts with compressed KV states and fetches the full Target KV Cache for verification. SpecOffload \cite{ref34} targets throughput-oriented inference when model weights cannot fit entirely in GPU memory and places Target weights and KV states across memory tiers. Dovetail \cite{ref35} assigns Target verification to the CPU and drafting to the GPU, coordinating heterogeneous execution under limited GPU resources.

\textbf{Parallel speculation and serving schedules.} General-purpose serving systems improve resource utilization through iteration-level batching \cite{ref36}, chunked prefill \cite{ref37}, and phase disaggregation \cite{ref38}. MineDraft \cite{ref10} overlaps drafting and verification through a pipeline across batches, SwiftSpec \cite{ref12} optimizes disaggregated execution and data exchange, PEARL \cite{ref39} overlaps Draft and Target execution via pre-verify and post-verify, and SPECTRE \cite{ref11} reuses remote Draft services. 


In contrast, SpecStream combines within-round multi-query sharing, chunked KV transfers, and online softmax, allowing full-attention Target verification to proceed before restoring the current layer's full KV history. It also exploits compute bubbles caused by Target KV-transfer waits to advance the Draft on the same GPUs, enabling parallel speculation without an additional Draft GPU.

\section*{Conclusion}

We present SpecStream, which jointly designs history management, streaming computation, and scheduling on shared GPUs around the state and access characteristics of the KV Cache in speculative decoding. The commit boundary separates history migration from candidate rollback. Queries within a round share the full history as it arrives in chunks, while Target-priority scheduling advances the Draft during data waits. Our experiments show that organizing KV access and model execution around the complete speculative decoding step can both alleviate GPU memory limits and improve serving efficiency under offloading. In future work, we plan to extend KV offloading beyond CPU memory to additional storage tiers.

\label{end:main}
\FloatBarrier
\normalsize


\begin{thebibliography}{42}
\setlength{\itemsep}{0pt}
\setlength{\parsep}{0pt}
\setlength{\parskip}{0pt}

\bibitem{ref40}
Tom B. Brown, Benjamin Mann, Nick Ryder et al. \href{\detokenize{https://arxiv.org/abs/2005.14165}}{Language Models are Few-Shot Learners}. Advances in Neural Information Processing Systems (NeurIPS 2020), vol. 33, pp. 1877--1901.

\bibitem{ref01}
Woosuk Kwon, Zhuohan Li, Siyuan Zhuang et al. \href{\detokenize{https://doi.org/10.1145/3600006.3613165}}{Efficient Memory Management for Large Language Model Serving with PagedAttention}. Proceedings of the 29th Symposium on Operating Systems Principles (SOSP 2023), pp. 611--626.

\bibitem{ref02}
Ying Sheng, Lianmin Zheng, Binhang Yuan et al. \href{\detokenize{https://proceedings.mlr.press/v202/sheng23a.html}}{FlexGen: High-Throughput Generative Inference of Large Language Models with a Single GPU}. Proceedings of the 40th International Conference on Machine Learning (ICML 2023), vol. 202, pp. 31094--31116.

\bibitem{ref03}
Yaniv Leviathan, Matan Kalman, Yossi Matias. \href{\detokenize{https://proceedings.mlr.press/v202/leviathan23a.html}}{Fast Inference from Transformers via Speculative Decoding}. Proceedings of the 40th International Conference on Machine Learning (ICML), 2023, vol. 202, pp. 19274--19286.

\bibitem{ref04}
Charlie Chen, Sebastian Borgeaud, Geoffrey Irving et al. \href{\detokenize{https://arxiv.org/abs/2302.01318}}{Accelerating Large Language Model Decoding with Speculative Sampling}. arXiv preprint arXiv:2302.01318, 2023.

\bibitem{ref06}
Yuhui Li, Fangyun Wei, Chao Zhang et al. \href{\detokenize{https://proceedings.mlr.press/v235/li24bt.html}}{EAGLE: Speculative Sampling Requires Rethinking Feature Uncertainty}. Proceedings of the 41st International Conference on Machine Learning (ICML), 2024, vol. 235, pp. 28935--28948.

\bibitem{ref05}
Tianle Cai, Yuhong Li, Zhengyang Geng et al. \href{\detokenize{https://proceedings.mlr.press/v235/cai24b.html}}{Medusa: Simple LLM Inference Acceleration Framework with Multiple Decoding Heads}. Proceedings of the 41st International Conference on Machine Learning (ICML), 2024, vol. 235, pp. 5209--5235.

\bibitem{ref07}
Xupeng Miao, Gabriele Oliaro, Zhihao Zhang et al. \href{\detokenize{https://doi.org/10.1145/3620666.3651335}}{SpecInfer: Accelerating Large Language Model Serving with Tree-based Speculative Inference and Verification}. Proceedings of the 29th ACM International Conference on Architectural Support for Programming Languages and Operating Systems (ASPLOS), 2024, vol. 3, pp. 932--949.

\bibitem{ref08}
Bin Gao, Zhuomin He, Puru Sharma et al. \href{\detokenize{https://www.usenix.org/conference/atc24/presentation/gao-bin-cost}}{Cost-Efficient Large Language Model Serving for Multi-turn Conversations with CachedAttention}. 2024 USENIX Annual Technical Conference (USENIX ATC 2024), pp. 111--126.

\bibitem{ref09}
Wonbeom Lee, Jungi Lee, Junghwan Seo et al. \href{\detokenize{https://www.usenix.org/conference/osdi24/presentation/lee}}{InfiniGen: Efficient Generative Inference of Large Language Models with Dynamic KV Cache Management}. 18th USENIX Symposium on Operating Systems Design and Implementation (OSDI 2024), pp. 155--172.

\bibitem{ref10}
Zhenwei Tang, Arun Verma, Zijian Zhou et al. \href{\detokenize{https://arxiv.org/abs/2603.18016}}{MineDraft: A Framework for Batch Parallel Speculative Decoding}. International Conference on Machine Learning (ICML), 2026. arXiv:2603.18016.

\bibitem{ref11}
Jincheng Xie, Yawen Ling, Qi Xiao et al. \href{\detokenize{https://arxiv.org/abs/2605.08151}}{SPECTRE: Hybrid Ordinary-Parallel Speculative Serving for Resource-Efficient LLM Inference}. arXiv preprint arXiv:2605.08151, 2026.

\bibitem{ref12}
Ziyi Zhang, Ziheng Jiang, Chengquan Jiang et al. \href{\detokenize{https://doi.org/10.1145/3779212.3790246}}{SwiftSpec: Disaggregated Speculative Decoding and Fused Kernels for Low-Latency LLM Inference}. Proceedings of the 31st ACM International Conference on Architectural Support for Programming Languages and Operating Systems (ASPLOS), 2026, vol. 2, pp. 2197--2211.

\bibitem{ref41}
An Yang, Anfeng Li, Baosong Yang et al. \href{\detokenize{https://arxiv.org/abs/2505.09388}}{Qwen3 Technical Report}. arXiv preprint arXiv:2505.09388, 2025.

\bibitem{ref42}
InternLM Team. \href{\detokenize{https://huggingface.co/internlm/internlm2_5-20b-chat}}{InternLM2.5-20B-Chat}. Official Hugging Face model card. Accessed 2026-08-25.

\bibitem{ref17}
Joshua Bakita, James H. Anderson. \href{\detokenize{https://www.cs.unc.edu/~jbakita/rtas23.pdf}}{Hardware Compute Partitioning on NVIDIA GPUs}. Proceedings of the 29th IEEE Real-Time and Embedded Technology and Applications Symposium (RTAS), 2023, pp. 54--66.

\bibitem{ref13}
Tri Dao, Daniel Y. Fu, Stefano Ermon et al. \href{\detokenize{https://proceedings.neurips.cc/paper/2022/hash/67d57c32e20fd0a7a302cb81d36e40d5-Abstract-Conference.html}}{FlashAttention: Fast and Memory-Efficient Exact Attention with IO-Awareness}. Advances in Neural Information Processing Systems (NeurIPS 2022), vol. 35, pp. 16344--16359.

\bibitem{ref14}
Tri Dao. \href{\detokenize{https://proceedings.iclr.cc/paper_files/paper/2024/hash/98ed250b203d1ac6b24bbcf263e3d4a7-Abstract-Conference.html}}{FlashAttention-2: Faster Attention with Better Parallelism and Work Partitioning}. The Twelfth International Conference on Learning Representations (ICLR 2024).

\bibitem{ref15}
NVIDIA. \href{\detokenize{https://docs.nvidia.com/deploy/mps/latest/index.html}}{Multi-Process Service}. NVIDIA CUDA Documentation. Accessed 2026-09-12.

\bibitem{ref16}
NVIDIA. \href{\detokenize{https://images.nvidia.com/aem-dam/en-zz/Solutions/data-center/nvidia-ampere-architecture-whitepaper.pdf}}{NVIDIA A100 Tensor Core GPU Architecture}. Architecture Whitepaper, 2020. Accessed 2026-09-12.

\bibitem{ref18}
Lianmin Zheng, Liangsheng Yin, Zhiqiang Xie et al. \href{\detokenize{https://proceedings.neurips.cc/paper_files/paper/2024/hash/724be4472168f31ba1c9ac630f15dec8-Abstract-Conference.html}}{SGLang: Efficient Execution of Structured Language Model Programs}. Advances in Neural Information Processing Systems (NeurIPS 2024), vol. 37, pp. 62557--62583.

\bibitem{ref19}
Karl Cobbe, Vineet Kosaraju, Mohammad Bavarian et al. \href{\detokenize{https://arxiv.org/abs/2110.14168}}{Training Verifiers to Solve Math Word Problems}. arXiv preprint arXiv:2110.14168, 2021.

\bibitem{ref20}
Yushi Bai, Shangqing Tu, Jiajie Zhang et al. \href{\detokenize{https://aclanthology.org/2025.acl-long.183/}}{LongBench v2: Towards Deeper Understanding and Reasoning on Realistic Long-context Multitasks}. Proceedings of the 63rd Annual Meeting of the Association for Computational Linguistics (Volume 1: Long Papers), 2025, vol. 1, pp. 3639--3664.

\bibitem{ref21}
OpenAI. \href{\detokenize{https://huggingface.co/datasets/openai/mrcr}}{OpenAI MRCR: Long context multiple needle in a haystack benchmark}. Hugging Face dataset card, 2025.

\bibitem{ref32}
Hanshi Sun, Zhuoming Chen, Xinyu Yang et al. \href{\detokenize{https://arxiv.org/abs/2404.11912}}{TriForce: Lossless Acceleration of Long Sequence Generation with Hierarchical Speculative Decoding}. First Conference on Language Modeling (COLM), 2024.

\bibitem{ref22}
Yuhui Li, Fangyun Wei, Chao Zhang et al. \href{\detokenize{https://aclanthology.org/2024.emnlp-main.422/}}{EAGLE-2: Faster Inference of Language Models with Dynamic Draft Trees}. Proceedings of the 2024 Conference on Empirical Methods in Natural Language Processing (EMNLP), pp. 7421--7432.

\bibitem{ref23}
Yuhui Li, Fangyun Wei, Chao Zhang, Hongyang Zhang. \href{\detokenize{https://proceedings.neurips.cc/paper_files/paper/2025/hash/c7b5a35ea98b62512a869c19ea7b03cb-Abstract-Conference.html}}{EAGLE-3: Scaling up Inference Acceleration of Large Language Models via Training-Time Test}. Advances in Neural Information Processing Systems (NeurIPS), 2025, vol. 38, pp. 136737--136756.

\bibitem{ref24}
Zhuoming Chen, Avner May, Ruslan Svirschevski et al. \href{\detokenize{https://proceedings.neurips.cc/paper_files/paper/2024/hash/ea1f5f0878d43ff4fb8bf64ef4a2326c-Abstract-Conference.html}}{Sequoia: Scalable and Robust Speculative Decoding}. Advances in Neural Information Processing Systems (NeurIPS), 2024, vol. 37, pp. 129531--129563.

\bibitem{ref25}
Ranajoy Sadhukhan, Jian Chen, Zhuoming Chen et al. \href{\detokenize{https://proceedings.iclr.cc/paper_files/paper/2025/hash/13f972adf12bdf886583d48cd528002f-Abstract-Conference.html}}{MagicDec: Breaking the Latency-Throughput Tradeoff for Long Context Generation with Speculative Decoding}. The Thirteenth International Conference on Learning Representations (ICLR), 2025.

\bibitem{ref26}
Rishabh Tiwari, Haocheng Xi, Aditya Tomar et al. \href{\detokenize{https://proceedings.mlr.press/v267/tiwari25b.html}}{QuantSpec: Self-Speculative Decoding with Hierarchical Quantized KV Cache}. Proceedings of the 42nd International Conference on Machine Learning (ICML), 2025, vol. 267, pp. 59668--59686.

\bibitem{ref27}
Jinwei Yao, Kaiqi Chen, Kexun Zhang et al. \href{\detokenize{https://arxiv.org/abs/2404.00242}}{DeFT: Decoding with Flash Tree-attention for Efficient Tree-structured LLM Inference}. The Thirteenth International Conference on Learning Representations (ICLR), 2025.

\bibitem{ref28}
Jiayi Yao, Hanchen Li, Yuhan Liu et al. \href{\detokenize{https://doi.org/10.1145/3689031.3696098}}{CacheBlend: Fast Large Language Model Serving for RAG with Cached Knowledge Fusion}. Proceedings of the Twentieth European Conference on Computer Systems (EuroSys 2025), pp. 94--109.

\bibitem{ref29}
Yang Liu, Yunfei Gu, Liqiang Zhang et al. \href{\detokenize{https://www.usenix.org/conference/fast26/presentation/liu-yang}}{CacheSlide: Unlocking Cross Position-Aware KV Cache Reuse for Accelerating LLM Serving}. 24th USENIX Conference on File and Storage Technologies (FAST 2026), pp. 83--99.

\bibitem{ref30}
Cheng Luo, Zefan Cai, Hanshi Sun et al. \href{\detokenize{https://arxiv.org/abs/2502.12574}}{HeadInfer: Memory-Efficient LLM Inference by Head-wise Offloading}. arXiv preprint arXiv:2502.12574, 2025.

\bibitem{ref31}
Shibo Jie, Yehui Tang, Kai Han et al. \href{\detokenize{https://proceedings.mlr.press/v267/jie25a.html}}{SpeCache: Speculative Key-Value Caching for Efficient Generation of LLMs}. Proceedings of the 42nd International Conference on Machine Learning (ICML), 2025, vol. 267, pp. 27917--27928.

\bibitem{ref33}
Jiayi Yao, Samuel Shen, Kuntai Du et al. \href{\detokenize{https://arxiv.org/abs/2605.17613}}{VeriCache: Turning Lossy KV Cache into Lossless LLM Inference}. arXiv preprint arXiv:2605.17613, 2026.

\bibitem{ref34}
Xiangwen Zhuge, Shen Xu, Zeyu Wang et al. \href{\detokenize{https://arxiv.org/abs/2505.10259}}{SpecOffload: Unlocking Latent GPU Capacity for LLM Inference on Resource-Constrained Devices}. arXiv preprint arXiv:2505.10259, 2025.

\bibitem{ref35}
Libo Zhang, Zhaoning Zhang, Baizhou Xu et al. \href{\detokenize{https://aclanthology.org/2025.emnlp-main.879/}}{Dovetail: A CPU/GPU Heterogeneous Speculative Decoding for LLM inference}. Proceedings of the 2025 Conference on Empirical Methods in Natural Language Processing (EMNLP), pp. 17382--17395.

\bibitem{ref36}
Gyeong-In Yu, Joo Seong Jeong, Geon-Woo Kim et al. \href{\detokenize{https://www.usenix.org/conference/osdi22/presentation/yu}}{Orca: A Distributed Serving System for Transformer-Based Generative Models}. 16th USENIX Symposium on Operating Systems Design and Implementation (OSDI 2022), pp. 521--538.

\bibitem{ref37}
Amey Agrawal, Nitin Kedia, Ashish Panwar et al. \href{\detokenize{https://www.usenix.org/conference/osdi24/presentation/agrawal}}{Taming Throughput-Latency Tradeoff in LLM Inference with Sarathi-Serve}. 18th USENIX Symposium on Operating Systems Design and Implementation (OSDI 2024), pp. 117--134.

\bibitem{ref38}
Yinmin Zhong, Shengyu Liu, Junda Chen et al. \href{\detokenize{https://www.usenix.org/conference/osdi24/presentation/zhong-yinmin}}{DistServe: Disaggregating Prefill and Decoding for Goodput-optimized Large Language Model Serving}. 18th USENIX Symposium on Operating Systems Design and Implementation (OSDI 2024), pp. 193--210.

\bibitem{ref39}
Tianyu Liu, Yun Li, Qitan Lv, Kai Liu, Jianchen Zhu, Winston Hu, and Xiao Sun. \href{\detokenize{https://proceedings.iclr.cc/paper_files/paper/2025/hash/03b1043052700b1a471996b0baf309d4-Abstract-Conference.html}}{PEARL: Parallel Speculative Decoding with Adaptive Draft Length}. The Thirteenth International Conference on Learning Representations (ICLR 2025).

\end{thebibliography}
\end{document}